\documentclass[preprint]{aastex}
\usepackage{color}

\begin{document} 

\title{Spectral and Imaging Observations of a White-light Flare\\ in the Mid-Infrared} 

\author{Matt Penn}
\affil{National Solar Observatory \footnote{NSO is operated by AURA Inc, under contract
to the National Science Foundation}, Tucson, AZ 85719}
\email{mpenn@nso.edu}

\author{S{\" a}m Krucker}
\affil{University of Applied Sciences and Arts Northwestern Switzerland, CH-5210 Windisch, Switzerland}
\affil{Space Sciences Laboratory, UC Berkeley}

\author{Hugh Hudson}
\affil{Space Sciences Laboratory, UC Berkeley}

\author{Murzy Jhabvala}
\affil{ Instrument Systems and Technology Division, Code 550, NASA Goddard Space Flight Center}

\author{Don Jennings}
\affil{Detector Systems Branch Code 553, NASA Goddard Space Flight Center}

\author{Allen Lunsford}
\affil{Department of Physics, the Catholic University of America}

\author{Pierre Kaufmann}
\affil{Center of Radio Astronomy and Astrophysics, Engineering School, \\ Mackenzie Presbyterian University, S{\~ a}o Paulo, Brazil}

\begin{abstract}
We report high-resolution observations at mid-infrared wavelengths of a minor solar flare, SOL2014-09-24T17T17:50 (C7.0), using Quantum Well Infrared Photodetector (QWIP) cameras at an auxiliary of the McMath-Pierce telescope.
The flare emissions, the first simultaneous observations in two mid-infrared bands at $5\ \mu$m and  $8\ \mu$m with white-light and hard X-ray coverage, revealed impulsive time variability with increases on time scales of $\sim 4$~s followed by exponential decay at $\sim$10 s in two bright regions separated by about 13$''$. The brightest source is compact, unresolved spatially at the diffraction limit ($1.3''$ at $5\ \mu$m).
We identify the IR sources as flare ribbons also seen in white-light emission at 6173~\AA~observed by SDO/HMI, with twin hard X-ray sources observed by RHESSI, and with EUV sources (e.g., 94~\AA) observed by SDO/AIA.
The two infrared points have closely the same flux density ($f_\nu$, W/m$^2$Hz) and extrapolate to a level about an order of magnitude below that observed in the visible band by HMI, but with a flux more than two orders of magnitude above the free-free continuum from the hot ($\sim$15 MK) coronal flare loop observed in the X-ray range.
The observations suggest that the IR emission is optically thin; this constraint and others suggest major contributions from a  density less than about $3 \times 10^{13}$~cm$^{-3}$.
We tentatively interpret this emission mechanism as predominantly free-free emission in a highly ionized but cool and rather dense chromospheric region.
\end{abstract}

\section*{Introduction}

Solar flares (and their analogs in many astrophysical contexts) consist of sudden energy releases, typically producing broad-band radiations mainly as the end result of the acceleration of particles to high energies. 
Even for the Sun this acceleration can lead to GeV particles and corresponding $\gamma$-ray continua, and of course the radio spectrum reflects a huge variety of phenomena as the solar corona (or stellar envelope) makes drastic adjustments to its magnetic field \citep[e.g.,][]{2011SSRv..159...19F}.
The observable broad-band spectrum of a solar flare, from kHz to GeV, typically has many gaps and uncertainties, but recently bolometric observations from space have begun to set definite limits on the energies involved \citep{2004GeoRL..3110802W}.
Heretofore missing and possibly very important spectral bands include the UV/EUV, including the Lyman continuum \citep{1978SoPh...59..129M} and the infrared-submm band \citep{1975SoPh...43..405O}.
We need the information provided in these bands to understand the flow of energy in the flare process; in particular the infrared-submm band can provide a vital link between the powerful photospheric emission of the white-light flare itself \citep[e.g.][]{1989SoPh..121..261N} and the dynamics of the upper atmosphere, including coronal mass ejections and related ``space weather.''

Earlier observations had revealed a flare continuum in the near infrared at 1.56~$\mu$ \citep{2004ApJ...607L.131X} and in the mid-infrared as well \citep{2013ApJ...768..134K,2015SoPh..tmp..141T}.
This exciting development allows us to extend our understanding of flare continuum emissions, known in white light to represent a dominant component of the luminous energy ever since the original observations by \cite{1859MNRAs..20...13C} and \cite{1859MNRAs..20...16H}.
In this article we describe a significant step forward in filling the observational gap in the mid-infrared spectral region (2--25$\mu$m) by the use of Quantum Well Infrared Photodetector (QWIP) 
cameras at the 0.81-m East Auxiliary branch of the McMath-Pierce telescope at the National Solar Observatory at Kitt Peak, Arizona, with a diffaction limit of 2.55$''$ at the centroid response wavelength of $8\ \mu$m, and 1.62$''$ at 5$\ \mu$m. The flare we report, SOL2014-10-24 (C7.0), also produced white-light emission detected by the Helioseismic and Magnetic Imager \citep{2012SoPh..275..207S} on the Solar Dynamics Observatory spacecraft, and there is complete coverage in the hard X-ray range by the Reuven Ramaty High Energy Solar Spectrocopic Imager \citep[RHESSI,][]{2002SoPh..210....3L}.

Our interpretation of SOL2014-09-24 generally agrees with that presented by \cite{2015SoPh..tmp..141T} for the long-duration flare event SOL2012-03-13 (M7.9), a white-light flare with $\gamma$-radiation, and with that presented by \cite{2015JGRA..120.4155K} for the flare SOL2014-10-27 (X2).
Trottet et al. place the bulk of the mid-IR source for SOL2012-03-13 at the top of the flare chromosphere, as identified in the semi-empirical model F2 of \cite{1980ApJ...242..336M}, with some contribution also from the photosphere.
Our flare differs substantially in its morphology, though: it also had white-light continuum but was highly impulsive and had no detectable $\gamma$-rays.
The more recent event SOL2014-10-27 also could be detected in two sub-THz bands \citep{2015JGRA..120.4155K} with a rough consistency in the positve-slope spectral energy distribution up to the mid-IR wavelengths.

\section*{Infrared Camera and Observations}

The McMath/Pierce East Auxiliary telescope is an open all-reflecting heliostat telescope with wavelength transmission from UV to IR wavelengths as determined by atmospheric transmission and an aperture of 0.76~m.  
The f/50 design produces a large solar image at prime focus.  
Mounted at this focus is a reimaging bench with two off-axis parabolic mirrors, which de-magnifies the image onto the 
Quantum Well Infrard Photodetector (QWIP) \citep{2007qwip.book.....S} at roughly 0.76 arcsec per pixel. 
This image scale was not selected to match the telescope resolution at these wavelengths (which is about 1.6 and 2.5 arcsec at 5 and 8~$\mu$m respectively) but rather to produce sufficient spatial coverage by the 320$\ \times\ $256-pixel detector to image a large active region, thus increasing the odds of observing a flare.  
While the QWIP detector has a low quantum efficiency, the large telescope aperture and large image scale produces a high photon flux at each pixel, which in turn dictates short camera exposures of less than 10 ms.

\begin{figure}[htbp]
\epsscale{0.75}
\plotone{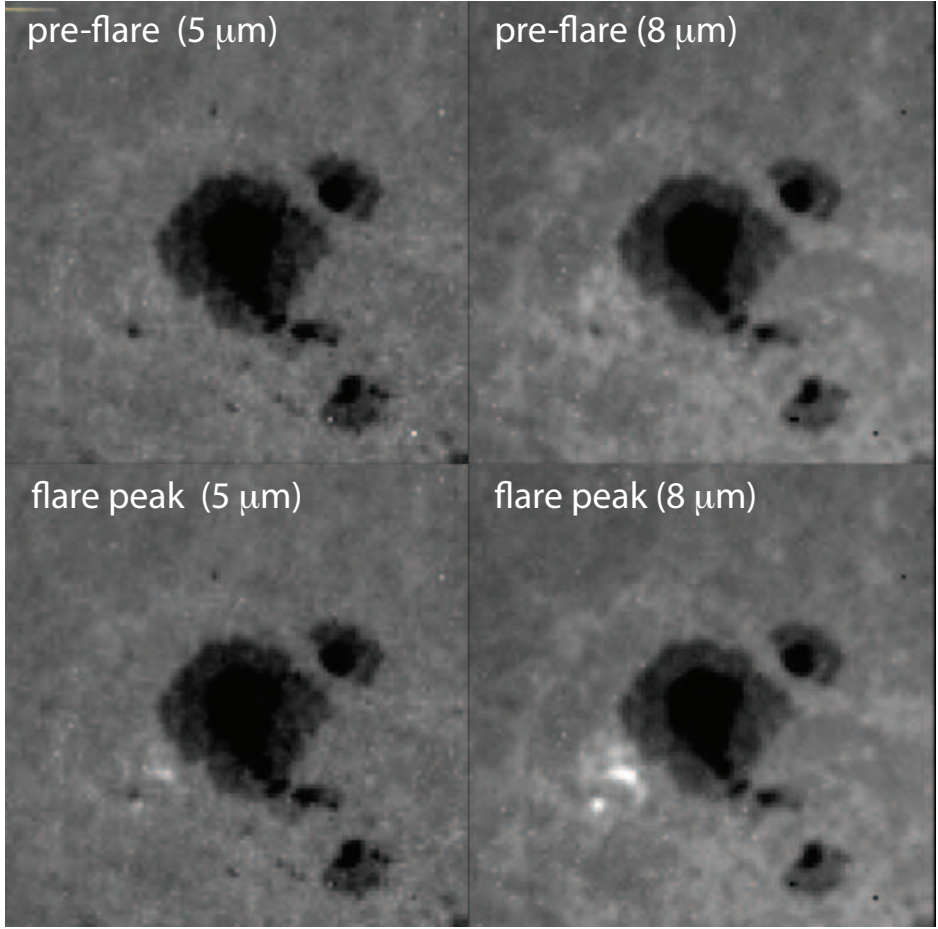}
\caption{Snapshot raw images: top row pre-flare, bottom row flare peak (17:49:20~UT), with 5~$\mu$m left and 8~$\mu$m right.}
\label{fig:penn_images}
\end{figure}

The QWIP detector used for this work is a two-color detector based on a GaAs substrate and built onto a ISC0006 silicon-readout integrated circuit.
The pixel  absorption bands are tuned to respond to excitation by photons of different energies, and the spectral response for the two colors reach 20\% of the maximum response at wavelengths of 4.2 to 6.2~$\mu$m for the 5~$\mu$m channel, and 7.0 to 9.3~$\mu$m for the 8~$\mu$m channel.  
Multiplying the wavelength response of the detector by the transmission of the Earth's atmosphere shifts the central wavelengths by small amounts, to 5.2 and 8.2~$\mu$m, for nominal observing conditions.
Because of uncertainty about the source spectrum and the exact atmospheric transmission at the time of observation, we approximate the effective response wavelengths to 5~ and 8~$\mu$m in this paper.
While the pixels in each channel sample exactly the same part of the image plane, they have slightly different exposure times which are different by a few milliseconds; at the cadence of these flare observations this is unimportant.  
The camera electronics can read and store images at a speed up to 15 frames per second, but the observations here have been recorded at only one frame per second.  
Dark and gain corrections were done for all of the images, and image de-rotation and alignments were also made.  
The data for this paper focus on a small spatial region of the FOV centered on the flare emission, for detailed analysis, and at 2$\times$2-pixel binning.  
Figure~\ref{fig:penn_images} shows images of this smaller region centered on the flare in both colors, taken before the flare and during the peak flare emission.

These images show well-defined sunspots and facular regions, as expected from earlier observations \citep[e.g.][]{1970SoPh...14..112T}.
The mid-IR continuum, though still formed in the low photosphere, begins the  transition in height from photospheric facula to chromospheric plage, and the properties of the bright non-flaring regions also hold considerable interest.
The infrared observations do not have absolute coordinate references, and so we have also used these features for empirical co-alignment to HMI and RHESSI images.

\begin{figure}[htbp]
\epsscale{1}
\plotone{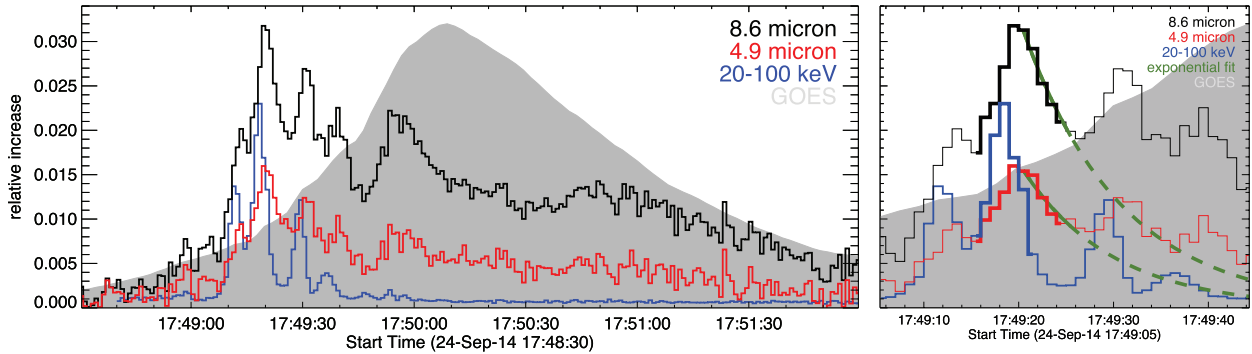}
\caption{Normalized time series ($\Delta S/\bar{S}$) for the compact southern sources of each wavelength, plus arbitrarily normalized time series from RHESSI hard X-rays (blue) and from the GOES 1-8~\AA~band (grey shaded).
The RHESSI data have been demodulated to get higher time resolution \citep[e.g.][]{2012A&A...547A..72Q}.
}
\label{fig:IR_figure2_time_long}
\end{figure}

Figure~\ref{fig:IR_figure2_time_long}  
shows the flare development in the two IR bands as relative increases above the pre-flare level, $\Delta S/\bar{S}$, where $\Delta S = S - \bar{S}$ is the increase of the flux $S$ above the pre-flare level $\bar{S}$. 
Note the excellent signal-to-noise ratio here and its improvement at the longer wavelength.
The rapid variability implies that the mid-IR emissions follow the non-thermal energy release of the impulsive phase, as already known independently from another event reported at 30~THz \citep[10~$\mu$m,][]{2013ApJ...768..134K}.
This and the appearance of two pimary emission patches strongly suggests that we can identify these impulsive mid-IR sources with flare footpoints, as commonly observed in hard X-rays and generally at many wavelengths, starting with the white-light continuum observed by Carrington in 1859.
The time variations at the two sources are closely simultaneous (Figure~\ref{fig:IR_figure2_time_long}), to within the 1~s sampling presented here.
We find an approximately $\sim$4~s rise time for the brightest peaks at 8~$\mu$m, followed by an exponential decay that is slower, with typical decay times of $\tau \sim 10$ s. 
The compact source and short duration hint that the observations approach the basic scales of energy release, but we emphasize that the spatial scale remains unresolved, implying that the inferred time scale should be treated as an upper limit.
During the decay phase of the soft X-ray emission, a longer-lasting IR emission also appears, but we do not analyze this slower component here. 
Unfortunately, the absolute timing of the IR emission for this event is uncertain due to a mishap in the clock synchronization. 
The IR lightcurves shown in the figures have been adjusted by eye, so that the rise phases in IR coincide with the HXR peaks as it would be the case if the HXR producing electrons heat chromospheric plasma. 
However, future observations must show if this approach is justifiable. 

\begin{figure}[htbp]
\epsscale{0.5}
\plotone{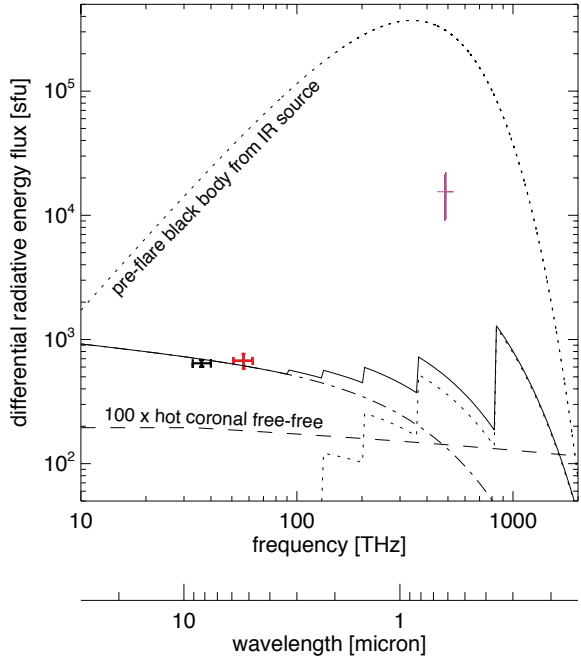}
\caption{Spectral energy distribution at flare peak for SOL2014-10-24 of the compact southern footpoint, covering the infrared-optical bands (black and red, corresponding to Figure~\ref{fig:IR_figure2_time_long}).
The magenta point shows the peak HMI signal (at 45-s integration, and so therefore to be interpreted as a lower limit).
The dotted line shows a blackbody at the photospheric effective temperature and normalized to the footpoint area of the southern source as defined for the IR signals in Figure~\ref{fig:IR_figure3_images}. 
The dashed line shows the free-free emission from the hot ($\sim$15 MK) coronal source derived from GOES at the peak time in the 1-8~\AA~channel.. 
This free-free emission, consistent with that expected from the hot flare loops, is much weaker than the IR footpoints and so the curve shown has been multiplied by 100 to fit on the plot.
The dotted and solid lines show free-free free-bound continua from an optically thin source at $1.5 \times 10^4$~K, adjusted to match the observations.
}
\label{fig:oir_spectrum}
\end{figure}

To calibrate the IR data, we normalize the quiet-Sun regions of the images to the known solar brightness temperatures at these wavelengths \citep[e.g.][]{1973asqu.book.....A}.
The relative increase shown in Figure~\ref{fig:IR_figure2_time_long} together with the area used in summing the imaging data can then be used to estimate the spectral flux (flux density). 
These fluxes at both IR wavelengths turn out to be the same within the error bars and the spectral shape appears to be flat (see Figure~\ref{fig:oir_spectrum}).
Following the discussion in \cite{2015SoPh..tmp..141T} of 30~THz (10~$\mu$m) observations of the M.9 flare SOL2012-03-13 \citep{2013ApJ...768..134K}, we identify the main contributor to this spectrum as the free-free component of recombination continuum from an optically thin, low-temperature plasma.

\section*{Hard X-rays and White Light}

\begin{figure}[htbp]
\epsscale{1}
\plotone{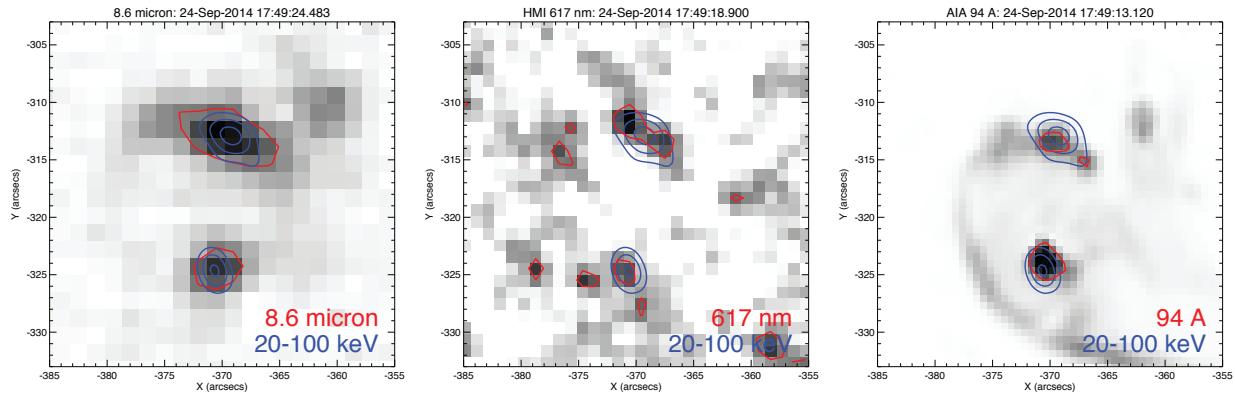}
\caption{Imaging of the flare ribbons in IR, HXR, and optical range: the image on the left shows the pre-flare subtracted $8\ \mu$m emission at the peak of the impulsive phase at 17:49:24~UT at a sub-second exposure time. Enhanced emission is shown in black. The blue contours give the 20-100 keV emissions at 50, 70, 90\%. from a RHESSI Clean image at 3.3$''$ FWHM resolution for a 32 second integration time around 17:49:24~UT. For reference, the red contours give the 50\% level of the $8~\mu$m emission. 
The middle panel shows the same for the 617 nm HMI image at 45 second resolution centered around 17:49:18.9UT,
and the right panel shows an AIA 94\AA~image (slightly earlier).}
\label{fig:IR_figure3_images}
\end{figure}

Hard X-ray observations from RHESSI also show twin sources, which we have empirically coaligned with the IR observations by direct correlation (Figure~\ref{fig:IR_figure3_images}). 
The excellent agreement between the IR and HXR sources (one compact, one somewhat extended) well justifies the co-alignment procedure. 
The HXR flux is unusually bright for a GOES class C7 flare, and the spectra are also unusually flat/hard around 30~keV. 
A standard thick-target fit gives a high low-energy cutoff of around 50~keV with an electron power law index of 3.8, resulting in an energy deposition rate of 1.1$\times$10$^{11}$ erg (cm$^2$ s)$^{-1}$ for electrons above 50 keV. 
In summary, the HXR observations show a rather uncommon non-thermal component, and the IR emission might therefore also be unusually bright compared to other flares. 
We expect future observations of other flares to clarify this point.

The flare, though only of GOES class C7, had a significant white-light signature as recorded by the SDO/HMI instrument in
its narrow passband near 617.3~nm.
As with the hard X-ray characteristics mentioned above, this also makes this flare somewhat distinctive.
Although the HMI excess images shows only a weak signal above the non-flare variations, the strongest white-light excesses matches the location of the HXR/IR source without the use of an empirical coalignment (Figure~\ref{fig:IR_figure3_images}). Furthermore, the elongated HXR/IR source appears also elongated. In white light  the flare can most easily be identified in a difference-image movie representation.
However the signal-to-noise ratio, on HMI's 45-s cadence, is not directly convincing for a base-difference time series.
Accordingly we have estimated the signal excess for this event by fitting a polynomial across the assumed flare interval (as judged by the cleaner IR signal) and using this as a background estimate. 
The peak fluxes of the flare at the three wavelengths (two IR, one visible) then allow us to plot the spectral energy distribution shown in Figure~\ref{fig:oir_spectrum}.

\section*{Analysis}

We assume optically-thin free-free radiation, based on the near equality of the flux densities at the two wavelengths.
Because the infrared wavelengths are beyond the H$^-$ recombination edge at 1.5~$\mu$m, this limits the recombination continuum to this component, for which
\begin{equation}\label{eq:tau}
\tau = 0.076 \times \frac{n_e n_i \ell}{\nu^2 T^{1.5}}\ \mathrm{cgs}.
\end{equation}

At longer wavelengths this flat spectrum must steepen to the Rayleigh-Jeans law, $f_\nu \propto \nu^2$.
The absence of a detected steepening sets a rough upper limit on the density within the source; assuming full ionization we find this to be at  $n_e \approx \ \times\ 10^{13}\ T_4^{3/4}\ l_7^{-1/2}$~cm$^{-3} $, where $T_4$ is the emission temperature in units of 10,000~K, taken here as $T\_4 = 1$, consistent with that estimated by \cite{2015SoPh..tmp..141T}.
This gives $n_e <  4 \times 10^{13}$~cm$^{-3}$ for $l_7$ = 0.5 (50~km), based on the VAL-C model \citep{1981ApJS...45..635V}.
Within the context of a standard semi-empirical model atmosphere, this would place the emission source in the mid-chromosphere or below, but the uncertainty of the opacity distribution makes this an approximate conclusion at best.
We do not expect that the major impulsive energy release necessary to create white-light flare emission would come close to the standard model assumptions of hydrostatic equilibrium, ionization balance, and one-dimensionality.

Another simple constraint on the physical conditions within the source comes from the observed flux, based on the assumption of optically-thin opacity in an isothermal slab (Equation~\ref{eq:tau}).
Figure~\ref{fig:oir_spectrum} shows a \textit{ad hoc} free-free continuum spectrum at 10$^4$~K, for which the emission measure is $7 \times 10^{50}$~cm$^{-3}$.
This leads to a density estimate $n_e \approx 4 \times 10^{13}$~cm$^{-3}$, assuming a source with the area corresponding to the 8~$\mu$m difraction limit of $1.5 \times 10^{16}$~cm$^{2}$ and $\ell_7 = 3$.
This rough estimate is consistent with the limit found above from the opacity constraint.

In principle other constraints on the physical conditions within the source could come from the observed time-series
variability, taking advantage of the high spatial and temporal resolution of the QWIP imaging.
In general flare emissions have gradual or impulsive components, with the latter associated with hard X-ray emission in dense chromospheric layers.
Here the cooling or dynamical time scales for newly-created hot sources can be short.
We do not discuss this quantitatively, but note that the observed time scales are consistent with typical white-light and HXR observations, both of which are currently limited by lack of spatial and temporal resolution, and remain essentially unresolved in both domains \citep{2006SoPh..234...79H}.

The infrared time-series photometry shows excellent signal-to-noise ratio, comparing flare emission excess with time-series fluctuations. We note that this is better, for this C7 event, than in white light for many X-class flares.
This implies a good expectation of success with observing other flares, and suggests that there might be continued improvement at still longer wavelengths, limited eventually by the poorer diffraction limit and probably by the approach to the Rayleigh-Jeans spectrum.

\section*{Conclusions}

This paper reports the first broad-band observations of the spectral energy distribution of a solar white-light flare in the mid-infrared ranges, with much-improved time resolution and signal-to-noise ratio.
We interpret the mid-IR continuum sources with the impulsive footpoint regions of the flare, as found previously by Trottet et al. (2015) for the mid-IR discovery event SOL2012-03-13 \citep{2013ApJ...768..134K}.
The observed timescales of the IR emission imply impulsive heating time scales less than $\sim$ 4~s, followed by exponential cooling with $\tau \sim$ 10~s. 
The observed two-point spectra in the mid-IR range are consistent with optical thin free-free bremsstrahlung emission from a slab.
To have optically thin emission, the density in the IR source needs to be lower than about $4 \times 10^{13}$~cm$^{-3}$. 
This constraint is consistent with an origin of the bulk of the mid-IR emission at the top of the flaring chromosphere, i.e.. immediately below the transition region.
The white-light flare matches as closely in space and in time as permitted by the observational limits, but must originate much deeper in the atmosphere.

The observed mid-IR/optical spectral energy distribution that we report leaves room for alternative interpretations.
In particular many ideas regarding the remarkable sub-THz component \citep{2004ApJ...603L.121K} of some flares include non-thermal emission mechanisms \citep[e.g.][]{2013A&ARv..21...58K}, and in the recent mid-IR event SOL2014-10-27 such sub-THz emission appeared in conjunction with mid-IR signatures similar to what we observe in the much weaker flare reported here.
Rigorously identifying the emission mechanisms will require more study, preferably with more complete spectral energy distributions. 
An additional QWIP camera which extends the spectral coverage to a third band at 13~$\mu$m was tested at the McMath-Pierce telescope in 2015. 
More observations are being made presently, typically following the daily guidance from flare probability forecasts (the �Max Millennium� target selection) to anticipate flare occurrence as well as possible. 
We wish to encourage analysis of these and other data thus obtained. 
Our success with the minor flare reported here suggests that the phenomenon of flare mid-IR emission may be commonplace, and as the analysis here has shown, we can learn substantial new things about flare structure from these wavelengths because of the excellent time resolution of the new data.

\begin{acknowledgements}
The development of the QWIP cameras was supported by the NASA Small
Business Innovative Research (SBIR) program, in collaboration with the
Goddard Space Flight Center and QmagiQ, LLC (Nashua, NH).
SK is supported by Swiss National Science Foundation (200021-140308) and through NASA contract NAS 5-98033 for RHESSI.
HSH thanks NASA for support under contract NNX11AP05.
The authors thank Paulo Sim{\~ o}es for correcting errors in the original draft.
\end{acknowledgements}

\bibliography{ir}

\bibliographystyle{apj}

\clearpage

\end{document}